\documentclass[epj,twocolumn]{webofc}
\usepackage[varg]{txfonts}   
\woctitle{Advances in Dark Matter and Particle Physics}
\begin{document}
\title{Excited lepton baryogenesis}
%
%

\author{Dmitry Zhuridov\inst{1}\fnsep\thanks{\email{dmitry.zhuridov@gmail.com}}
%
}

\institute{Institute of Physics, University of Silesia, Uniwersytecka 4, 40-007, Katowice, Poland 
          }

\abstract{%
  The excited leptons that share the quantum numbers with the 
Standard Model leptons but have larger masses are widespread in
many promising new physics theories. A subclass of excited leptons
that at low energies interact with the SM fermions dominantly
through the effective coupling to lepton and fermion-antifermion pair
can be referred as leptomesons. I introduce possible generation of
the baryon asymmetry of the universe using these new particles. The
discussed baryogenesis mechanisms do not contradict to the small neutrino
masses and the proton stability, and can be interesting for the
collider experiments.
}
\maketitle
\section{Introduction}
\label{seq-intro}
In spite of the present 
success of the Standard Model (SM) several observations indicate its possible nonfundamentality: 
large number of the SM fermions, their arbitrary masses and mixings, fractional electric charge of quarks, etc. 
Among the diversity of new physics models the theories of compositeness~\cite{Pati:1974yy,Terazawa:1976xx,Shupe:1979fv,
Squires:1980cm,Harari:1980ez,Fritzsch:1981zh,Barbieri:1981cy} try to solve these problems by introducing
a substructure of the SM particles, which 
subcomponents are commonly referred as preons~\cite{Pati:1974yy}.
%
The composite models may include in the particle content radial, orbital, topological, structural, 
and other excitations of the ground state particles, 
e.g., an {\it excited lepton} 
that shares leptonic quantum number with one of the existing leptons, has larger mass and no color charge. 

Besides the outlined issues on the 
particles and their interactions that come from the laboratory studies, 
another 
opened questions (including the dark matter problem) arise from the astrophysical observations 
of the universe around us. 
In particular, our universe appears to be populated exclusively with baryonic matter rather than 
\mbox{antimatter}
~\cite{PDG2016}. However this baryon asymmetry 
can not be explained within the Big Bang cosmology and the SM. Possible scenarios of dynamical generation of 
the baryon asymmetry during the evolution of the universe from a hot early 
matter-antimatter symmetric stage are referred as the baryogenesis (BG) mechanisms~\cite{Dolgov:1991fr,Dine:2003ax}, 
and include new physics.

In the next section we discuss one example of the composite models in question. 
The interactions and the mass bounds for the excited leptons are outlined in section~\ref{sec-Ex-leptons},
and the new BG scenarios that involve these particles are discussed in section~\ref{sec-BG}. Finally, 
we conclude in section~\ref{sec-conclusion}.

\section{Composite model example}
Consider the haplon models~\cite{Fritzsch:1981zh,Fritzsch:2014gta}, 
which are similar to the models with wakems and chroms~\cite{Terazawa:1979pj,Terazawa:1984dj,Terazawa:2011ci}, 
and are based on the symmetry ${\rm SU}(3)_c\times {\rm U}(1)_{em}\times {\rm SU}(N)_h$, 
where the new haplon group ${\rm SU}(N)_h$ has the confinement scale of the order of 
0.3~TeV, and denotes, e.g., ${\rm SU}(2)_L\times {\rm SU}(2)_R$. 
These models contain the two categories of preons (haplons): the fermions 
$\alpha^{+1/2}$ and $\beta^{-1/2}$, and the scalars $\ell^{+1/2}$ and $c_k^{-1/6}$, where $k=r,g,b$ 
(``red, green, blue'').
Their quantum number assignment is given in Table~\ref{tab-1}, where $Q$ is the electric charge, C1 is the choice for the 
${\rm SU}(3)_c$ representations in Ref.~\cite{Fritzsch:1981zh}\footnote{Notice that C1 assignment does not provide 
a spin-charge separation~\cite{Xiong:2016fum}.} and C2 is an alternative choice~\cite{Fritzsch:2014gta}. 
Then the haplon pairs can compose the SM particles as 
$\nu=(\alpha\bar\ell)$, $e^-=(\beta\bar\ell)$, $u=(\alpha\bar c_k)$, $d=(\beta\bar c_k)$, 
$W^-=(\bar\alpha\beta)$, $W^3=(\bar\alpha\alpha-\bar\beta\beta)/\sqrt{2}$,\dots, 
and the new particles, e.g., a scalar leptoquark $S^{+2/3}_\ell=(\ell\bar c_k)$, 
and the neutral scalars $S^0_\ell=(\ell\bar\ell)$ and $S^0_c=(c_k\bar c_k)$. 
$W^3$ mixes with the photon $\gamma$ similarly to the mixing between $\gamma$ and $\rho^0$-meson. 
$H$ scalar can be a p-wave excitation of the $Z$, and 
the second and third generations can be dynamical excitations. 
Notice that $S_\ell^{+2/3}$, $S_c^0$ and $S_\ell^0$ states (if their masses are small) may contribute to the low-energy observables, 
e.g., so-called, XYZ states~\cite{Ji:2016ffn}.
\footnote{
Notice that the SM results can be reproduced in some composite models, at least at the tree level, 
due to the complementarity between Higgs phase and confining phase~\cite{'tHooft:1998pk,Calmet:2000th}. 
}

\begin{table}
\centering
\caption{The haplon quantum numbers}
\label{tab-1}       
\begin{tabular}{llllll}
  \hline
  Haplon & Spin [$\hslash$] & $Q$ [$|e|$] & C1 & C2 & ${\rm SU}(2)_h$ \\
  \hline
  $\alpha$ & $1/2$ & $+1/2$ & 3 & 1 & 2 \\
  $\beta$ & $1/2$ &  $-1/2$ & 3 & 1 & 2 \\
  $\ell$ & 0 &  $+1/2$ & $\bar3$ & 1 & 2 \\
  $c_k$ & 0 & $-1/6$ & 3 & $3$ & 2 \\
  \hline
\end{tabular}
\end{table}

However, new questions arise: Where does this peculiar haplon picture come from? Can it, in turn, result from 
a substructure of haplons? 
Consider the two scalar ``prepreons''\footnote{For supersymmetric models with ``prepreons'' see, e.g., Ref.~\cite{Pati:1983dh}.} 
$\pi_k$ and $\bar\pi_k$, which are ${\rm SU}(3)$ triplets and have the electric charges 
of $-1/6$ and $+1/6$, respectively. 
Then the set of haplons with their electric and C2 color charges can be reproduced by the triples of ``prepreons'' 
($\bar\pi_{\bar r}\bar\pi_{\bar g}\bar\pi_{\bar b}\to\{\alpha,\ell\}$, $\pi_r\pi_g\pi_b\to\{\beta,\bar\ell\}$, 
$\bar\pi_{\bar i}\,\pi_j\pi_l\to c_k$), 
while additional mechanism of spin generation is required. 
One can think of possible relation of spin to 
a circular color currents similarly to some discussions in the context of 
the condensed matter~\cite{chuu_chang_niu} and gravity~\cite{Burinskii:2016agf} theories, 
taking into account that the distribution of matter in a composite state can be imagined (in particular, 
in the ${\rm SU}(3)$ Yang-Mills theory) 
in terms of the wave functions or probability distributions for the effective subcomponents of a finite 
size~\cite{Glazek:2016vkl,Glazek:2011wf,Gomez-Rocha:2015esa}. 
Then a ``spinning'' and ``nonspinning'' states of the same preon (e.g., $\alpha$ and $\ell$) may form 
a supersymmetric multiplet.

Notice that the possibility of multihaplon 
states such as $(\beta\bar c_k\bar\ell c_k)$, $(\alpha\bar\ell\beta\bar c_k\bar\beta c_k)$, etc., 
gets more points from recent discoveries of the multiquark hadrons~\cite{Aaij:2015tga}. 

\section{Excited leptons}
\label{sec-Ex-leptons}
The excited lepton states defined in the introduction can be particularly important if their masses are smaller than 
the leptoquark and leptogluon masses, which can be natural due to the absence of the color charge. 
The contact interactions among the SM fermions $f$ and the excited fermions $f^*$ can be generically written as~\cite{PDG2016}
\begin{eqnarray}\label{eq:CI}
      &&\mathcal{L}_\text{CI} = \frac{g_*^2}{2\Lambda^2} 
      \sum_{\alpha,\beta=L,R}  \left[  \eta_{\alpha\beta} (\bar f_\alpha\gamma^\mu f_\alpha)(\bar f_\beta\gamma_\mu f_\beta) 
       \right. \nonumber\\
      &&+ \left.\eta_{\alpha\beta}^\prime (\bar f_\alpha\gamma^\mu f_\alpha)(\bar f_\beta^*\gamma_\mu f_\beta^*)
      +  \tilde\eta_{\alpha\beta}^{\prime} (\bar f^*_\alpha\gamma^\mu f^*_\alpha)(\bar f_\beta^*\gamma_\mu f^*_\beta)
       \right. \nonumber\\
      &&+ \left.\eta_{\alpha\beta}^{\prime\prime} (\bar f_\alpha\gamma^\mu f_\alpha)(\bar f_\beta^*\gamma_\mu f_\beta)  
      +  \text{H.c.} + \dots \right],  
\end{eqnarray}
where $\Lambda$ is the contact interaction scale, $g_*^2=4\pi$, and the new parameter values are usually taken of $|\eta^j|\leq1$. 

Assuming nearly maximal couplings of $|\eta^j|\simeq1$ and the excited fermion masses of $M_{f^*}\simeq\Lambda$, 
the present lower bounds for $\Lambda/\sqrt{|\eta^j|}$ ratios are of the order of 
few~TeV~\cite{PDG2016}. 
However, if Eq.~\eqref{eq:CI} expresses a ``residue'' effective interactions between the composites (with respect to 
the fundamental interactions among their subcomponents) then $|\eta^j|$ couplings can be small, 
and even the case of $M_{f^*}\lesssim\Lambda<1$~TeV is not excluded for $|\eta^j|\ll1$ and $\Lambda/\sqrt{|\eta^j|}>>M_{f^*}$.

A particular type of excited leptons that at low energies interact with the SM fermions dominantly through 
the contact terms we refer as {\it leptomesons} (LM).\footnote{Notice that the same term ``leptomeson'' 
was used in the literature for the bound states of colored excitations of $e^+$ and $e^-$~\cite{Pitkanen:1992}.} 
The relevant contact terms (with $\eta^{\prime\prime}$ couplings) can be realized, e.g., through the leptoquark exchange.

\begin{figure*}
\centering
 \includegraphics[width=0.26\textwidth]{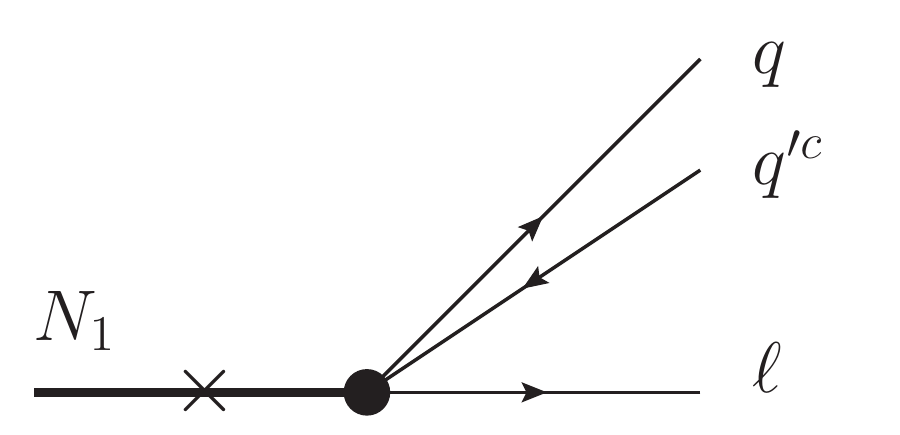} \
 \includegraphics[width=0.32\textwidth]{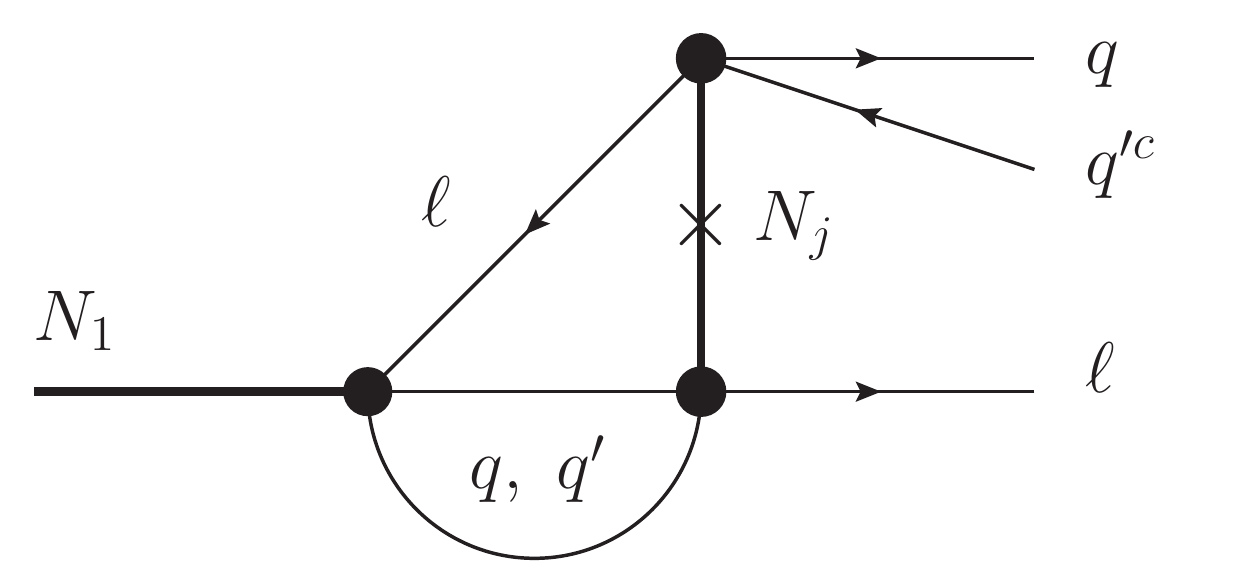} \,  
 \includegraphics[width=0.37\textwidth]{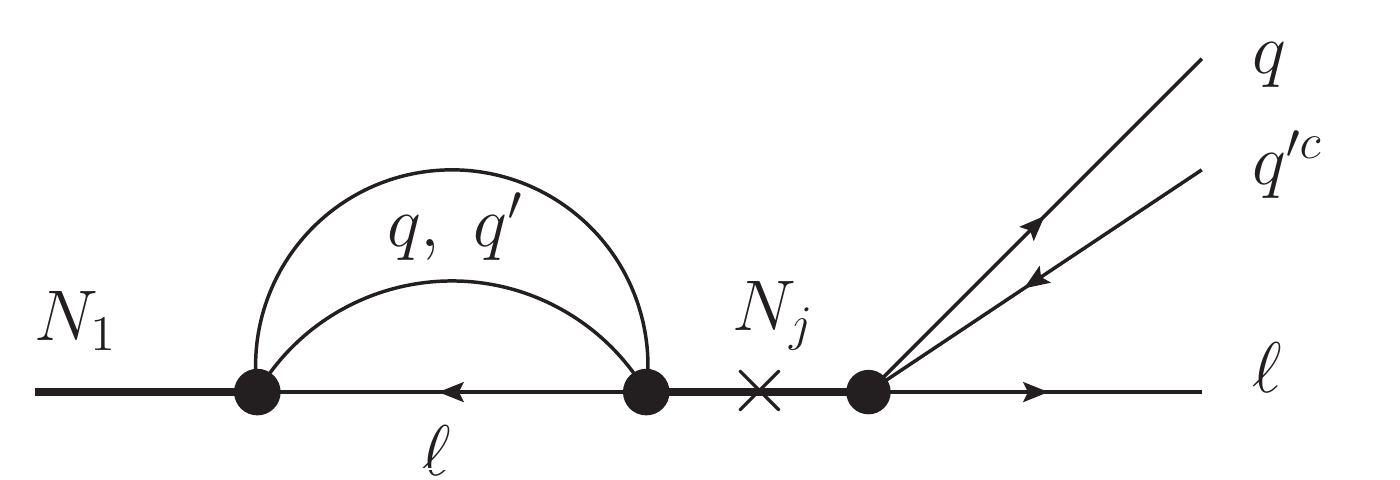}
\caption{Feynman diagrams for the discussed contributions to the CP asymmetry, where  
	   the line direction shows either $L$ or 
	   $B$ flow, ``$\times$'' represents a Majorana mass insertion, 
	   and the black bulb represents a subprocess (e.g., a leptoquark exchange).} 
   \label{Fig:LG_diagrams}
 \end{figure*}
\section{Baryogenesis}
\label{sec-BG}
Possible BG and the dark matter generation by a scalar \mbox{4-haplon} state was considered in Ref.~\cite{Fritzsch:2014gta}. 
In this proceedings we discuss if fermionic LM states can provide a successful BG~\cite{Zhuridov:2016xls}. 
Similarly to the sterile neutrino $\nu_R$ case, depending on the LM properties, deviation from thermal equilibrium can occur 
at either production or freeze-out and decay (compare to the BG via $\nu_R$ oscillations~\cite{Akhmedov:1998qx} and 
the usual leptogenesis~\cite{Fukugita_Yanagida}, respectively). 
%
In both scenarios one should replace the Yukawa interactions of $\nu_R$ by the contact interactions of LMs, 
which may result in promising effects. 

\subsection{BG from LM oscillations}
\label{sec-BG-LM-osc}
Once created in the early universe neutral long-lived LMs oscillate and interact with ordinary matter. 
These processes do not violate the total lepton number $L^\text{tot}$ (for Dirac LMs). 
However the oscillations violate $CP$ and therefore do not conserve individual lepton numbers $L_i$ for LMs. 
Hence the initial state with all zero lepton numbers evolves into a state with $L^\text{tot}=L+\sum_i L_i=0$ but $L_i\neq0$.

At temperatures $T\ll \Lambda$ the LMs communicate 
their lepton asymmetry to neutrinos $\nu_\ell$ and charged leptons $\ell$ 
through the effective interactions,
e.g., $B$-conserving (and $L$-conserving for Dirac LMs) vector couplings
\begin{eqnarray}\label{eq:4vertex}
    \sum_{\psi_\ell,f,f^\prime} 
    \sum_{\alpha,\beta=L,R}   \left[ 
    \frac{\epsilon^{\alpha\beta}_{ff^\prime\psi_\ell}}{\Lambda^2} 
    (\bar f_\alpha\gamma^\mu f_\alpha^\prime)(\bar\psi_{\ell\beta}\gamma_\mu N_{\ell\beta})   \right.\nonumber\\
  + \left.\frac{\tilde\epsilon^{\alpha\beta}_{ff^\prime\psi_\ell}}{\Lambda^2} 
    (\bar\psi_{\ell\alpha}\gamma^\mu f_\alpha^\prime)(\bar f_{\beta}\gamma_\mu N_{\ell\alpha})
    \right] 
    + \text{H.c.}, 
\end{eqnarray}
where $\psi_\ell=\ell,\nu_\ell$ ($\ell=e,\mu,\tau$), 
the constants 
$\stackrel{\scriptscriptstyle{(\sim)}}\epsilon=4\pi\eta^{\prime\prime}$ 
can be real, 
$f$ and $f^\prime$ denote either quarks or leptons 
such that 
$Q_{f_\alpha}+Q_{f^{\prime c}_\alpha}+Q_{\psi_{\ell\beta}}=0$, 
and $N_\ell$ is the neutral LM flavor state related to the mass eigenstates $N_i$ as 
$ N_{\ell\alpha} = \sum_{i=1}^n U^\alpha_{\ell i} N_i$.

Suppose that LMs of at least one type $N_i$ remain in 
thermal equilibrium till the electroweak symmetry breaking 
time $t_\text{EW}$ at which sphalerons become ineffective, and those of 
at least one other type $N_j$ come out-of-equilibrium by $t_\text{EW}$. 
Hence the lepton number of former (later) affects (has no effect on) the BG. 
In result, the final baryon asymmetry after $t_\text{EW}$ 
is nonzero. 
At the time $t\gg t_\text{EW}$ all LMs decay into the leptons and the quarks (hadrons).  
For this reason they do not contribute to the dark matter in the universe, and do not destroy 
the Big Bang nucleosynthesis. 

The system of $n$ types of singlet LMs of a given momentum $k(t)\propto T(t)$ that interact 
with the primordial plasma can be described by the $n\times n$ density matrix $\rho(t)$. 
In a simplified picture it 
satisfies the kinetic equation~\cite{Akhmedov:1998qx}
\begin{eqnarray}\label{eq:evolution}
  i \frac{d\rho}{dt} = [\hat H,\rho] - \frac{i}{2}\{\Gamma,\rho\} + \frac{i}{2}\{\Gamma^p,1-\rho\}, 
\end{eqnarray}
where $\Gamma$ ($\Gamma^p$) is the destruction (production) rate, and the effective Hamiltonian can be written as 
\begin{eqnarray}
  \hat H = V(t) + U \frac{\hat M^2}{2k(t)} U^\dag, 
\end{eqnarray}
where $\hat M^2=\text{diag}(M_1^2,\dots,M_n^2)$ with the masses $M_i$ of $N_i$, and $V$ is a real potential. 


One of the main features of the discussed BG from LMs is that
the 4-particle interaction cross section that contributes to the destruction rate is 
proportional to the total energy of the process $s$ instead of the inverse proportionality 
that takes place in the BG from $\nu_R$ 
oscillations. Indeed, this cross section can be written as
\begin{eqnarray}\label{eq:cross_section}
  \sigma(a+b\to c+d) = 
        \frac{C}{4\pi}|\epsilon|^2  \frac{s}{\Lambda^4}\propto s, 
\end{eqnarray}
where 
$a, b, c$ and $d$ denote the four interacting particles 
($f$, $f^\prime$, $\psi_\ell$ and $N_\ell$), and $C=\mathcal{O}(1)$ is the constant 
that includes the color factor in the case of the interaction with quarks. 
%
%
In result, the interaction rate that equilibrates LMs, 
\begin{eqnarray}\label{eq:rate}
  \Gamma \propto   |\epsilon|^2  \frac{T^5}{\Lambda^4}, 
\end{eqnarray}
is suppressed by 
$(T/\Lambda)^4$ with respect to the Higgs mediated interaction rate 
in usual BG via $\nu_R$ oscillations.

The conditions that LMs of type $N_i$ remain in thermal equilibrium till $t_\text{EW}$, 
while $N_j$ do not, can be written as
\begin{eqnarray}
  \Gamma_i(T_\text{EW}) >  H(T_\text{EW}),  \qquad 
  \Gamma_j(T_\text{EW}) <  H(T_\text{EW}),  
\end{eqnarray}
where $H(T)$ is the Hubble expansion rate. Due to the
suppression factor of $(T_\text{EW}/\Lambda)^4$ 
the successful BG can be realized with the relatively large couplings $|\epsilon|$ with respect to 
the sterile neutrino Yukawas of $Y\sim10^{-7}$ in the BG via $\nu_R$ oscillations~\cite{Akhmedov:1998qx}. 
In particular, for $\Lambda\gtrsim10$ and 30~TeV we have $|\epsilon|\gtrsim10^{-4}$ and $10^{-3}$, respectively. 
Hence the considered BG scenario can be relevant for the LHC and next colliders without unnatural hierarchy of the couplings.

\subsection{BG from LM decays}
\label{sec-BG-LM-dec}
Suppose that the neutral LMs are Majorana particles ($N=N^c$). 
Consider their out-of-equilibrium, $CP$- and $L$-violating decays 
in the early universe. 
The relevant interactions can be written as
\begin{eqnarray}
    \frac{\epsilon_{ff^\prime\psi_\ell}^{\alpha R}}{\Lambda^2} (\bar f_\alpha\gamma^\mu f_\alpha^\prime) 
    (\bar\psi_{\ell R}\gamma_\mu N_{\ell R}) + 
    \frac{\epsilon_{ff^\prime\psi_\ell}^{S}}{\Lambda^2} (\bar f_R f_L^\prime) 
    (\bar\psi_{\ell L} N_{\ell R}
    )  \nonumber\\
   +
    \frac{\epsilon_{ff^\prime\psi_\ell}^{T}}{\Lambda^2} (\bar f \sigma^{\mu\nu} f^\prime) 
    (\bar\psi_{\ell L} \sigma_{\mu\nu} N_{\ell R}
    ) + \text{H.c.} \hspace{1cm}  
\end{eqnarray}
To be more specific in the following we consider the term 
\begin{eqnarray}
    \frac{\lambda_{\ell i}}{\Lambda^2} (\bar q_\alpha\gamma^\mu q_\alpha^\prime) 
    (\bar\ell_R\gamma_\mu N_{iR}
    ),  
\end{eqnarray}
where $\lambda_{\ell i}=\epsilon_{qq^\prime\ell}^{\alpha R} U^R_{\ell i}$ is a complex parameter. 
Consider the interference of tree and one-loop diagrams\footnote{Same two-loop self-energy graph 
as in Fig.~\ref{Fig:LG_diagrams} (right) was 
discussed in the resonant BG scenarios of Refs.~\cite{Dev:2015uca,Davoudiasl:2015jja}, where the baryon asymmetry is 
directly produced in the three-body decays of the sterile neutrinos. 
Although these mechanisms involve $B$-violating interactions of $qqq\nu_R$ type, they do not lead to fast proton decay 
due to the large values of $\nu_R$ mass and the $B$-violating interaction scale of $\mathcal{O}(1)$~TeV. 
} 
shown in Fig.~\ref{Fig:LG_diagrams}. The final $CP$ asymmetry that is produced in decays of the lightest LMs $N_1$ 
\begin{eqnarray}
 \varepsilon_1 = \frac{1}{\Gamma_1}\, \sum_\ell [\Gamma(N_1 \to \ell_R q_\alpha q_\alpha^{\prime c}) 
 - \Gamma(N_1 \to \ell_R^c q_\alpha^c q_\alpha^\prime)], 
\end{eqnarray}
can be non-zero if $\text{Im}[ (\lambda^\dag \lambda)_{1j}^2 ] \neq 0$. 
Using the width~\cite{Cakir:2002eu},
\begin{eqnarray}
 \Gamma_1 &=& 
 {\sum_\ell[ \Gamma(N_1 \to \ell_R q_\alpha q_\alpha^{\prime c}) 
 + \Gamma(N_1 \to \ell_R^c q_\alpha^c q_\alpha^\prime)]}   \nonumber\\
 &\simeq&    \frac{1}{128\pi^3}\, (\lambda^\dag \lambda)_{11} \frac{M_1^5}{\Lambda^4}, 
\end{eqnarray}
the condition for the decay parameter $K\equiv\Gamma_1 / H(M_1) > 3$ 
(strong washout regime) translates into the limit of
\begin{eqnarray}
 (\lambda^\dag \lambda)_{11} \gtrsim   4 \times 10^{-7} \times \left( \frac{\Lambda}{10~\text{TeV}} \right)^4 \times
 \left( \frac{1~\text{TeV}}{M_1} \right)^3. 
\end{eqnarray}

The final baryon asymmetry can be written as 
\begin{eqnarray}
  \frac{n_B-n_{\bar B}}{s} = 
  \left(-\frac{28}{79}\right)\times \frac{n_L-n_{\bar L}}{s}  
         = \left(-\frac{28}{79}\right)\times \frac{\varepsilon_1 \kappa}{g_*},  
\end{eqnarray}
where $n_B$ ($n_L$) is the baryon (lepton) number density, $s$ is the entropy density, 
$-28/79$ is the sphaleron lepton-to-baryon factor, and $\kappa\leq1$ is the washout coefficient that can be determined by solving 
the set of Boltzmann equations. 
The observed baryon asymmetry of the universe of~\cite{PDG2016}
\begin{eqnarray}\label{eq:eta_B}
  \eta_B = \frac{n_B-n_{\bar B}}{n_\gamma} = 7.04 \times\frac{n_B-n_{\bar B}}{s} \simeq 6
  \times10^{-10},
\end{eqnarray}
where $n_\gamma$ is the photon number density, 
can be produced, e.g., for 
$K\sim100$ with the degeneracy factor of
\begin{eqnarray}
  \mu \equiv \frac{M_2-M_1}{M_1} \lesssim 10^{-6} \left(\frac{M_1}{1~\text{TeV}}\right), 
\end{eqnarray}
which enters the resonant $CP$ asymmetry of
\begin{eqnarray}
  \varepsilon_i \sim  \frac{\text{Im}\{[(\lambda^\dag\lambda)_{ij}]^2\}}{(\lambda^\dag\lambda)_{ii}
  (\lambda^\dag\lambda)_{jj}}   \frac{\Gamma_j}{M_j}   \frac{M_iM_j}{M_i^2-M_j^2}  
  \sim \mu^{-1} \frac{\Gamma_1}{M_1}.
\end{eqnarray} 

Notice that the discussed effective LM-quark-quark-lepton vertex can be realized, e.g.,  
through the exchange of ${\rm SU}(2)_L$ singlet scalar leptoquark $S_{0R}$ with 
$Y=1/3$. 
Relevant interaction terms can be written as
\begin{eqnarray}
    -\mathcal{L}_\text{int} &=& (g_{ij}\, \bar d_{R}^c N_{iR} 
	+  f_j\, \bar u^c_R \ell_R ) S_{0R}^j 
	+ \text{H.c.} 
\end{eqnarray}
Then the above expressions are valid with the replacements of 
$\lambda \to gf^*$ and $\Lambda \to M_{S_{0R}}$. 
The relevant to the successful BG 
values of the new couplings 
of $|g|\sim |f|\sim 0.01-0.1$, can be interesting for the collider searches.

\begin{figure}
\centering
 \includegraphics[width=4.5cm,clip]{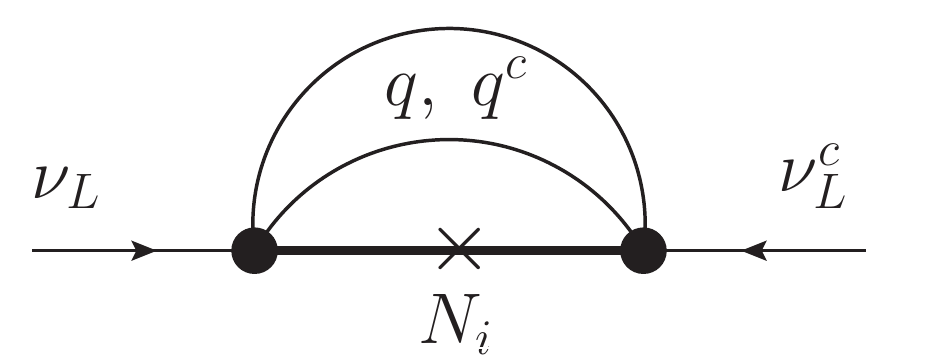}
\caption{The discussed contribution to the neutrino masses (line direction shows $L$ flow).}
   \label{Fig:nu_mass}
 \end{figure}
\begin{figure}
\centering
 \includegraphics[width=6.5cm,clip]{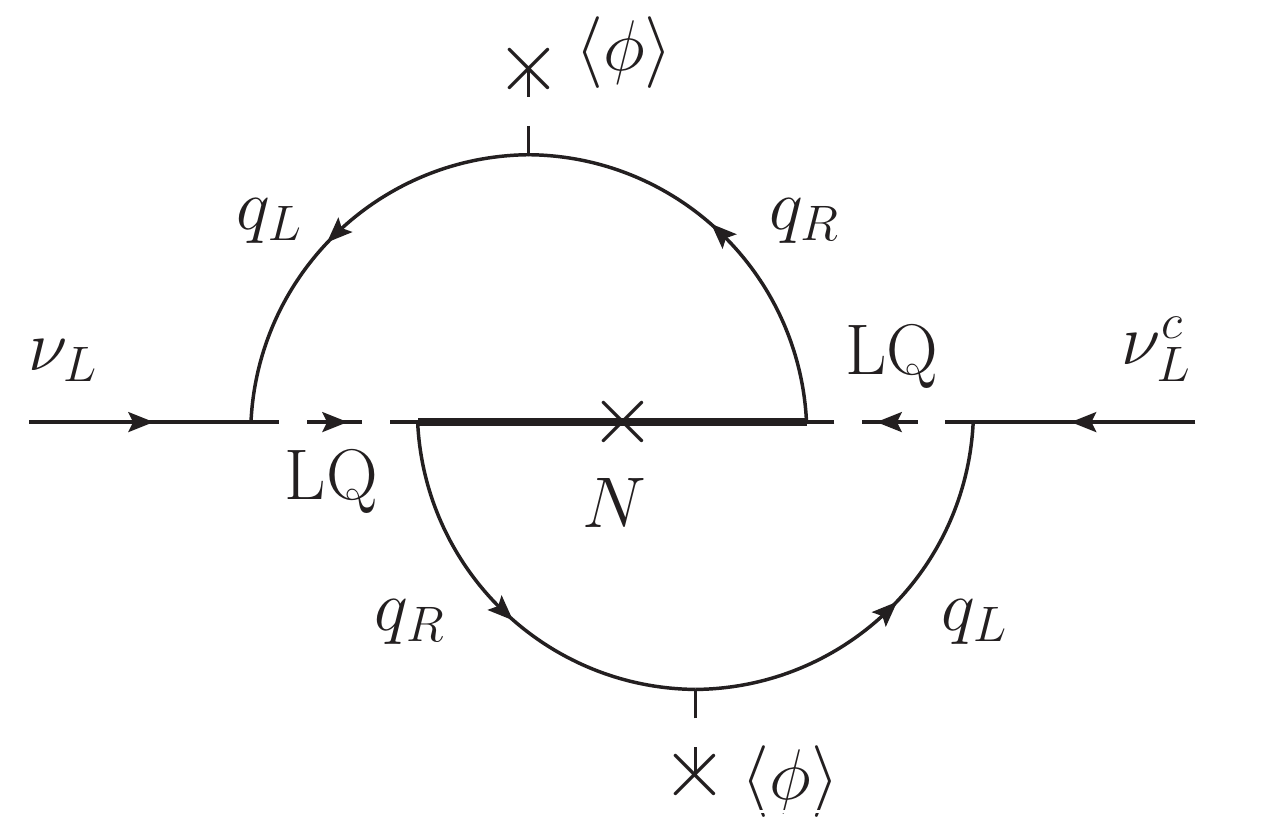}
\caption{Leptoquark (LQ) contribution to the neutrino masses.}
   \label{Fig:nu_mass_LQ}
 \end{figure}
\subsection{Neutrino masses}
\label{sec-nu-mass}
For Majorana LMs the effective terms of
\begin{eqnarray}
%
      \frac{\epsilon_{ff\nu_\ell}^{S}}{\Lambda^2} \bar f_R f_L \,
    \bar\nu_{\ell L} N_{\ell R} 
    +  \frac{\epsilon_{ff\nu_\ell}^{T}}{\Lambda^2} \bar f\sigma^{\mu\nu} f \,
    \bar\nu_{\ell L} \sigma_{\mu\nu} N_{\ell R}
    +  \text{H.c.}  
\end{eqnarray}
can generate the two-loop contributions to the observable light neutrino masses $m_{\nu_\ell}$ 
that is illustrated for $f=q$ in Fig.~\ref{Fig:nu_mass}, and for a particular model with the leptoquarks 
in Fig.~\ref{Fig:nu_mass_LQ}.
A simple estimate of this contribution is
\begin{eqnarray}
  m_{\nu_\ell} \sim  \sum_i \frac{(\epsilon\, U_{\ell i}^R)^2}{(16\pi^2)^2}\ \frac{M_i^3 m_q^2}{\Lambda^4},
\end{eqnarray}
where $m_q$ is the quark mass. 
Hence the present bound of $m_\nu\lesssim2$~eV~\cite{Aseev:2011dq} can be easily satisfied.

\section{Conclusion}
\label{sec-conclusion}

The two new testable baryogenesis scenarios in the models with excited leptons are
introduced, which do not contradict to the observable neutrino masses. 
First, {\it the BG from LM oscillations} may take place for relatively light and long-lived LMs, 
which do not all decay before $t_\text{EW}$. 
Second, {\it the BG from LM decays} can be realized if all LMs decay before $t_\text{EW}$.
A particular models based on the former (later) BG proposal require detailed study of the neutrino potential 
(of the Boltzmann equations) to be verified in the future experiments.
In both scenarios the baryon number is violated only by the sphaleron processes 
that does not affect the proton stability.
Due to the relatively low temperatures of the discussed BG mechanisms, 
an analog of the gravitino problem~\cite{Khlopov1,Balestra} does not exist there.

\section*{Acknowledgements} 
The author thanks Marek Zra{\l}ek, Henryk Czy\.z, Bhupal Dev 
and Yue Zhang for useful discussions. 
This work was supported in part by the Polish National Science Centre, grant number DEC-2012/07/B/ST2/03867. 
The author used JaxoDraw~\cite{Binosi:2003yf} to draw the Feynman diagrams.

\end{document}